# We cite as we communicate:
# A communication model for the citation process

Victor V. Kryssanov[a,*], Evgeny L. Kuleshov[b], Frank J. Rinaldo[a], and Hitoshi Ogawa[a]

[a] Faculty of Information Science and Engineering, Ritsumeikan University, Kusatsu, Shiga 525-8577, Japan
[b] Institute of Physics and Information Technologies, the Far-Eastern National University, Vladivostok 690950, Russia

**Abstract**
Building on ideas from linguistics, psychology, and social sciences about the possible mechanisms of human decision-making, we propose a novel theoretical framework for the citation analysis. Given the existing trend to investigate citation statistics in the context of various forms of power and Zipfian laws, we show that the popular models of citation have poor predictive ability and can hardly provide for an adequate explanation of the observed behavior of the empirical data. An alternative model is then derived, using the apparatus of statistical mechanics. The model is applied to approximate the citation frequencies of scientific articles from two large collections, and it demonstrates a predictive potential much superior to the one of any of the citation models known to the authors from the literature. Some analytical properties of the developed model are discussed, and conclusions are drawn. Directions for future work are also given at the paper's end.

## 1. Introduction

Over the past two decades, there is a growing body of publications in both special and popular media discussing the otherwise familiar phenomenon of social recognition of professional achievements of those engaged in scientific activities at either individual or institutional levels. Apart from somewhat mercantile yet natural interest from the scientific community, there are at least two major factors lying behind this growth: the rocketing cost of research and development in most of the scholarly disciplines and the increasing transparency of the decision-making policies and procedures of various resource-distributing organizations. There is, therefore, a continuous search for fair, optimal and universal ways to evaluate and possibly reward particular contributions in the global knowledge development process. Citation analysis constitutes a significant, if not the largest part of the research efforts undertaken in this direction.

Citation analysis builds on statistics reflecting the dynamics of referring to relevant and contingently important work by authors of scientific and technical papers, and it aims at unveiling the social mechanisms responsible for the development of knowledge in a research community. Two high-profile articles by D. J. S. Price published in 1965 are often considered a pioneering attempt to explain the observed (irr)regularities in knowledge proliferation, as it is represented in academic writing [1,2]. It may be said that it was Price who triggered a substantial interest to the topic from the general public and, at the same time, prompted specialists in different fields, ranging from theoretical physics to sociology, to come forward with possible explanations of the citation trends and practices. In 1968, R. K. Merton suggested a "Matthew effect" model by explicitly reconstructing a "cumulative advantage" algorithm (see Ref. [3]) to account for the citation process [4]. It was projected that the social status – popularity, esteem, merit, or else "usefulness" – of one or another work can be explicated in terms of simple behavioral configurations prevailing among the members of the community to which the work is reported. Merton, perhaps after Price, thus conjectured that the citers' choice, based on priorities, might be the cause of the "scale-free" patterns detected in citation statistics.

---
[*]Corresponding author.
*E-mail addresses:* kvvictor@is.ritsumei.ac.jp (V.V. Kryssanov), kuleshov@lemoi.phys.dvgu.ru (E.L. Kuleshov), rinaldo@is.ritsumei.ac.jp (F.J. Rinaldo), ogawa@airlab.ics.ritsumei.ac.jp (H. Ogawa).

Preferential choice (later re-coined as "preferential attachment" by Barabasi and Albert [5]) became the cornerstone idea for social network theories, which emerged over the years to explain the apparently scale-free behavior peculiar to citation as well as to many other social and natural phenomena typically discussed in the context of various power and Zipfian laws [6,7]. Quite paradoxically, however, there exists little empirical support for but mounting evidence against this model along with its numerous cousins. Even large collections of citation data do not exhibit scale-free properties in the whole data range, as citation frequencies never follow a single straight line – the signature of the power law – when plotted on a double-logarithmic scale. Furthermore, the existing practice of the citation model "validation," where the principal argument is built around whether a histogram of the artificial data resembles, as to an eye, the one of the actual data, can hardly be considered irrefutable [8,9]. The poor predictive performance – statistical unsoundness – of the models makes highly questionable any speculations about the possible "social meaning" and/or indicative ability of the model parameters. The same should, unfortunately, be said about a handful of alternative models developed in the past 10 years and advocated to replace the "canonical" power law with a sum of two Pareto distributions [10], a lognormal [11], stretched exponential [12], or rather exotic modified Bessel [13], or Tsallis [14,15] distribution to mimic citation occurrence data (see Fig. 1). (In Fig. 1 and throughout the text of this article, the term CCDF will be used to denote a complementary cumulative distribution function, in the continuous case, and a complementary cumulative sum, in the case of discrete random variables. For a random variable $Z$, CCDF is formally defined as $1-F(z)$, where $F(z)$ represents the distribution function $F(z) = \Pr[Z \leq z]$, $\Pr[A]$ denotes the probability of event $A$ occurring.)

Regardless of the statistical significance testing, which however alone may preclude a practical use other than marginal academic deliberations of any theory, the very idea of reducing the complex machinery of human decision-making to some sort of behavioral ordering, such as preferential choice, although appealing in its simplicity, appears unfounded and lacks empirical support. The computational modeling experiments conducted to demonstrate the emergence of the power law out of individual interactions have no evident links with relevant studies in cognitive and experimental psychology and sociology, hence "differential diagnostics" and across-domain checking of the modeling approaches is hardly possible, if ever discussed at all (for a related debate, see Ref. [16]; also – Ref. [17]). On the other hand, by looking into the detailed structure of citation data, it has been discovered that, depending on the discipline and country of origin, up to 99% of academic reports may never be cited (the uncitedness rate ranges from 36 to 88% across different fields, and from 9 to 99% – across sub-disciplines[18,19]), some 50% are not read [20,21], and from 47 to 75% of citations to scholarly manuscripts come from authors outside the field [22]. At the same time, over 80% of citations in a single paper may well be copied from other publications [23], from 4 to 17% are self-citations [24], and about 7% are erroneous [25]. These facts seriously undermine the feasibility of introspective analysis of human decision-making in the citation process and, taking into account the arguments of the above paragraphs, force us to conclude that the existing understanding of the citation phenomenon is unsatisfactory, while its popular theoretical models are wrong by any standard recognizable in the contemporary science.

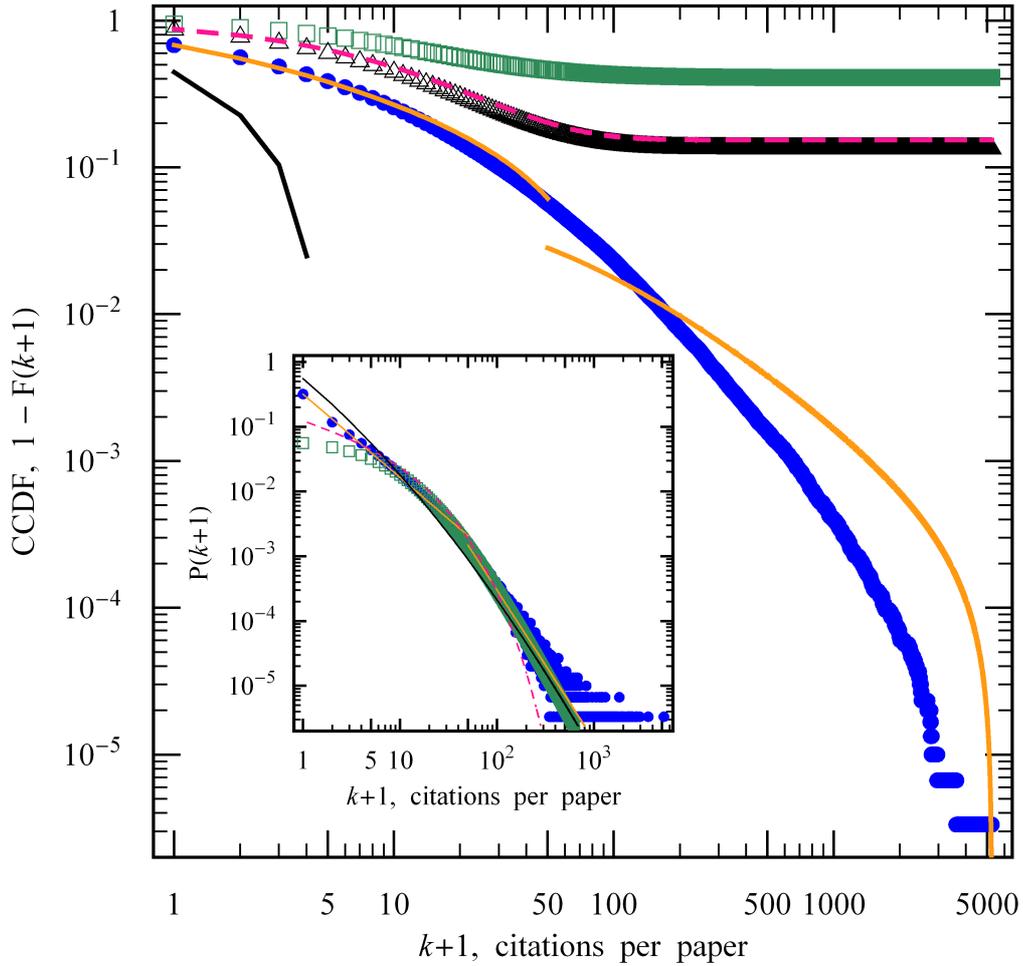

**Fig. 1.** Models of the citation process: A data sample (shown with dots) comprising citation information of about 300 000 papers in high energy physics (for details on the data, see Section 3) was fitted to five models (for details on the models, consult the references listed at the end of the 3$^{rd}$ paragraph, Section 1; in the formulas, P($k$) denotes the probability of $k$, a citation rate): a linear combination of two power laws (light solid lines; for each segment, P($k$) $\sim k^{-a}$, $a$ is a parameter, estimated using a non-linear MSE method; "$\sim$" denotes "equal up to a multiplicative constant"), a lognormal form (dark solid line; P($k$) $\sim e^{-b\ln k - c(\ln k)^2}$, $b$, $c$ – parameters, by minimizing a chi-square fit statistic), a stretched exponential (light dashed line; P($k$) $\sim (k/a)^{b-1} e^{-(k/a)^b}$, $a$, $b$ – parameters, by numerical MLE), a modified Bessel (triangles; P($k$) $\sim 2a\, \mathbf{I_0}(2\sqrt{ak})$, $\mathbf{I_0}(\cdot)$ denotes a modified Bessel function of the 0$^{th}$ order, and $a$ is a parameter, by numerical MLE), and a Tsallis (squares; P($k$) $\sim [1+(q-1)\lambda k]^{-\frac{q}{q-1}}$, $q$ and $\lambda$ are parameters, estimated by minimizing a chi-square fit statistic in an iterative algorithm). None of the models provides an adequate, which is to say statistically significant, fit.
Inset: The same results presented in an obviously misleading but overwhelmingly popular manner with the corresponding histograms (to improve visibility, the modified Bessel model is not displayed, as it appears similar to the stretched exponential fit).

The objective of this work is to explore the citation process as a communication phenomenon. The underlying motivation is twofold: firstly, to demonstrate that citation can and, in fact, should be analyzed in a broad context of the communication studies that would include relevant results from neurophysiology, psychology, linguistics, and behavioral science, and secondly, to derive a mathematical model that would withstand a thorough statistical testing and give accurate predictions of citation statistics.

## 2. Theory

The main premise of our modeling approach is that citation is an information exchange process that does not differ in principle from other modes of communication and that uses traditional communication strategies. The theoretical considerations of this paper arose naturally out of the generalization of a model of text- and hypertext-based communication reported elsewhere [26].

The citation process can be viewed as a mechanism for substantiating important, as they appear to the author, ideas – concepts, problems, statements, and the like – of a paper. Citations create the foundation upon which the paper arguments are based, assertions are asserted, and assumptions are left unexplained. It is always up to the reader to fully explore the foundations of the new idea or concept present in a given text, just as the same reader may or may not read the main text in full. We will, therefore, consider that every instance of referring to an external text (or, in other words, external communication) indicates the act of signification (by the citation) of a concept or idea otherwise not presented (e.g. owing to space limitations or legal reasons) in sufficient detail in the main text with other media, such as words, formulas, diagrams, etc. Every citation is then to be taken as a "meta-word" that may or may not be in the reader's dictionary and yet may not bear the same implied meaning for all concerned. Citations are part of a language the authors use to communicate their work to the public. We will further assume that the citing results from cognitive processing (problem solving or decision-making) associated with the signified idea. The processing time $\tau$ – the time for which the idea attracts attention (e.g. as in the case of a recognizably important problem), either personal or collective – depends on many factors, such as complexity of the underlying concept, social status of its originator, current "fashion" in the field, results obtained in related domains, etc. but generally exceeds the time-frame allocated for writing a particular paper.

For a statistical ensemble of citers, let us consider $K_0$ a discrete random variable indicating the count of different citations utilized to represent one fixed concept (no matter how complex). It appears natural to assume that values of $K_0$ will depend on the concept processing time $\tau$: for any fixed observation time, the longer, on average, the processing time, the more frequent the citation (an article with zero citations corresponds to zero processing time, e.g. when the considered concept/problem is trivial, outdated, notorious, "uninteresting," remains unnoticed, or is still poorly understood by anybody other than the article authors). The latter effectively means that there is a positive correlation between a number of citations that a specific problem receives and the time for which this problem remains in the focus. As in the case of words, the same citation may stand for different concepts, and the same concept may be signified with different citations.

To characterize the behavior of $K_0$, we will seek to estimate $f_{K_0}(s)$, $s = 1, 2, ...$, its probability mass function, PMF. (For technical reasons, we define the domain of the distribution as strictly positive, i.e. $s > 0$; the non-inclusion of the $0^{th}$ frequency is, however, not intrinsic to the method and is unnecessary when the model is derived in the continuous domain, e.g. see Ref. [27].) Building on the arguments by E. T. Jaynes [28], who showed that the least biased, "most true" estimate possible on the given information is, in many cases, the distribution that maximizes the Shannon entropy $H = -\sum_s f_{K_0}(s) \ln f_{K_0}(s)$, $f_{K_0}(s)$ can be obtained by maximizing under the normalization $\sum_s f_{K_0}(s) = 1$ and expectation $\sum_s s f_{K_0}(s) = \bar{k}_0$ constraints the following functional:

$$\mathcal{L}(f_{K_0}) = -\sum_s f_{K_0}(s) \ln f_{K_0}(s) + \gamma(1 - \sum_s f_{K_0}(s)) + \beta(\bar{k}_0 - \sum_s s f_{K_0}(s)), \qquad (1)$$

where $\bar{k}_0$ specifies the expectation of $K_0$, and $\gamma$ and $\beta$ are Lagrangian multipliers. From optimality conditions $\partial \mathcal{L}/\partial f_{K_0}(s) = 0$ and $\partial \mathcal{L}/\partial \gamma = 0$, one can easily derive:

$$f_{K_0}(s) = \Pr[K_0 = s \mid \beta] = (e^{\beta} - 1) e^{-s\beta}, \quad s = 1, 2, \ldots . \tag{2}$$

It is important to note that since $\bar{k}_0 = \sum_{s=1}^{\infty} s f_{K_0}(s) = e^{\beta}/(e^{\beta} - 1) = 1/f_{K_0}(1)$, for $\bar{k}_0 \gg 1$, $0 < f_{K_0}(1) \ll 1$ and $0 < \beta \ll 1$ that stipulates $\beta \approx 1/\bar{k}_0$. Hence, $1/\beta$, $\beta > 0$, may be used as an estimate of the citation average relative (i.e. for one concept) rate.

The form (2) can be interpreted as the most likely distribution – a distribution realized experimentally in overpoweringly more ways than any other candidate distribution, provided that the citation average relative rate is fixed [29]. The bursty nature of human decision-making dictates, however, that there should be considered two independent processes with different citation averages: somewhat low rate $1/\beta_1$ – for works not cited before (or at least recently), and a higher rate $1/\beta_2$ – for works that have just been cited. The corresponding PMF is written as follows:

$$g_{K_0}(s) = c(e^{\beta_1} - 1) e^{-s\beta_1} + (1 - c)(e^{\beta_2} - 1) e^{-s\beta_2}, \quad s = 1, 2, \ldots, \tag{3}$$

where $0 < c \leq 1$ gives the "weight" of the citation "first occurrence" (sub)process. While many other purposive interpretation arguments, such as individual vs. social appreciation, semantic (content) and syntactic (function) constituents of the communication, or (yes!) a "rich get richer" logic, could be brought forward to justify the introduction of the more general form (3) for the "one concept" citations, we will rely on the, perhaps least speculative, assertion that new problems (and their solutions) are reported at a substantially lower rate compared to problems that have already been discussed in the literature for a while. In terms of $\tau$, the latter would mean that the time expended for the problem solution (consideration, discussion, analysis, etc.) is affected by two independent processes, where the first is associated with (recently) unreported, while the second – with publicized ideas (concepts, theories, and the like).

We will now consider $K$ a measured stochastic variable indicating citation occurrences for not just one but many and different concepts. The statistical properties of $K$ depend on the parameters of the distribution (3), which can naturally vary (e.g. as a result of a variation in the processing time among different concepts and/or domains). $\beta_1$ and $\beta_2$ are thus to be defined as positive, continuous, and independent random variables; let $f(\beta_1)$ and $f(\beta_2)$ be their respective probability density functions (PDF). When the number of concepts/problems signified with citations is sufficiently large, $\mathrm{P}(k)$, $k = 1, 2, \ldots$, the PMF of $K$ can be obtained as

$$\mathrm{P}(k) = \int_0^{\infty} \int_0^{\infty} g_{K_0}(k) f(\beta_1) f(\beta_2) d\beta_1 d\beta_2 . \tag{4}$$

To arrive at a functional form of $\mathrm{P}(k)$, we now need to identify the distributions of $\beta_1$ and $\beta_2$. Utilizing the earlier made assumption that the citation rate is determined by the dynamics of problem-solving, it can be written that $f(\beta_1) \stackrel{d}{=} f(1/\tau_1)$ and $f(\beta_2) \stackrel{d}{=} f(1/\tau_2)$, where $\tau_1$ and $\tau_2$ are the times, for which first- and repeatedly- reported, in that order, problems receive (i.e. are represented with) citations, and "$\stackrel{d}{=}$" stands for "distributionally equal." Among a variety of distributions used in psychology and social sciences to mimic human decision-making time, the Inverse Gaussian (also sometimes called Wald) distribution is often regarded as the best-established model (see Ref. [30] for a thorough discussion of the relevant distributions; also – Ref. [31]). In the decision-making context, the Inverse Gaussian distribution can be obtained from a sequential sampling evidence accrual model as the distribution of first hitting times $T = \inf[\tau > 0, V(\tau) \geq b]$ to an absorbing barrier $b > 0$ of a space ($X$) and time homogeneous Wiener diffusion process $dV = \alpha\, d\tau + \sigma\, dx$ with $V(0) = 0$ initial condition [32]:

$$f(\tau \mid b, \alpha, \sigma) = \frac{b}{\sigma\sqrt{2\pi\tau^3}} e^{-\frac{(b-\alpha\tau)^2}{2\sigma^2\tau}}, \qquad (5)$$

where $\alpha > 0$ is the drift rate, and $\sigma > 0$ is a diffusion constant. Without loss of generality, the barrier can be fixed by setting $b = 1$. In terms of $\mu = 1/\alpha$, which is the so-called integration time (a characteristic of human information processing usually used as an indicator of the solved task difficulty), $\lambda = 1/\sigma^2$ an estimate of the non-randomness (sometimes called "precision," which would alternatively be interpreted by analogy with the physical diffusion as "activation effort"), and $\beta = 1/\tau$, we, after simple algebra, arrive at the following PDF:

$$f(\beta) = \sqrt{\frac{\lambda}{2\pi\beta}}\, e^{-\frac{\lambda(\beta\mu - 1)^2}{2\beta\mu^2}}, \quad \beta > 0, \qquad (6)$$

with the expectation and variance specified, correspondingly, as $E(\beta) = 1/\mu + 1/\lambda$ and $\mathrm{Var}(\beta) = (2\mu + \lambda)/(\mu\lambda^2)$.

The direct substitution of the form (6) for the random variables $\beta_1$ and $\beta_2$ into Eq. (4) yields:

$$P(k) = c\, e^{\lambda_1/\mu_1} \sqrt{\lambda_1} \left( \frac{e^{-\sqrt{\frac{\lambda_1(2k-2+\lambda_1)}{\mu_1^2}}}}{\sqrt{2k-2+\lambda_1}} - \frac{e^{-\sqrt{\frac{\lambda_1(2k+\lambda_1)}{\mu_1^2}}}}{\sqrt{2k+\lambda_1}} \right) + (1-c)\, e^{\lambda_2/\mu_2} \sqrt{\lambda_2} \left( \frac{e^{-\sqrt{\frac{\lambda_2(2k-2+\lambda_2)}{\mu_2^2}}}}{\sqrt{2k-2+\lambda_2}} - \frac{e^{-\sqrt{\frac{\lambda_2(2k+\lambda_2)}{\mu_2^2}}}}{\sqrt{2k+\lambda_2}} \right) \qquad (7)$$

that is thus the probability mass function of the citation occurrence number.

## 3. Experiment

To explore the appropriateness of the assumptions of the proposed theoretical framework and test the derived model (7) against empirical data, we have conducted an experiment using two datasets. The first collection contains citation statistics of $24\,296$ articles published in Physical Review D in 1975-1994 and cited at least once; the maximum number of citations received by a single paper is $2\,026$, and the total number of citations is $351\,872$. The data was obtained from http://physics.bu.edu/~redner/projects/citation/prd.html (last accessed on March 9, 2007), where it is publicly available. The second collection is comprised of citation data for $299\,239$ published journal articles, both cited and uncited, in high energy physics accumulated in the SLAC SPIRES database since 1962. The maximum citation number is $5\,242$, and the total number is $4\,024\,332$. This latter data was obtained from Sune Lehmann (for more details on the data, consult the work by Lehmann, Lautrup, and Jackson [10]), and it was also used to examine the alternative models shown in Fig. 1. In both collections, the data represent the professional activities of comparatively small research communities and may thus be thought of as reasonably homogeneous.

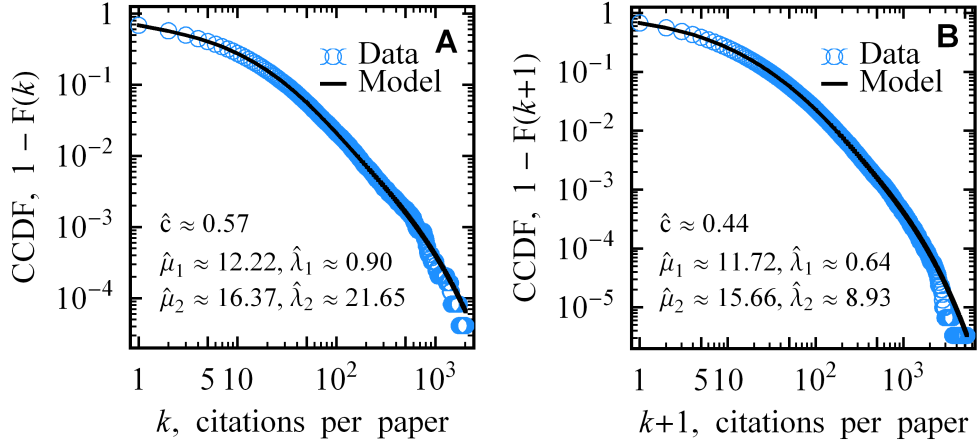

**Fig. 2.** Results of the modeling of citation statistics for articles from the Physical Review D collection (**A**) and the SLAC SPIRES dataset (**B**).

Fig. 2 displays results of the modeling of the observed citation frequencies with Eq. (7) (also see Figs 4, **B** and **C**). $\hat{c}$, $\hat{\mu}_1$, $\hat{\lambda}_1$, $\hat{\mu}_2$, and $\hat{\lambda}_2$ estimates for the corresponding parameters of the PMF were calculated using a numerical maximum likelihood method. In both cases, Pearson's $\chi^2$ test does not reject the proposed model with a significance level $\alpha = 0.1$. (To ensure the validity of the chi-square test, bins naturally formed by the discrete data but containing less than 5 elements have been merged.)

## 4. Discussion

The analytical framework presented in Section 2 provides for a remarkably good fit to the available data – a quality not shared by other citation mechanism models known to the authors from high-profile periodicals. Arguments about an excessive complexity of the derived 5-parameter distribution, although expected, are naturally defeated by the fact that empirical citation data reveal two distinct citation regimes, where each of the regimes forms a visibly curved line with citation frequencies plotted on a double-logarithmic scale (see Fig. 1, inset, and Fig. 4; also – Refs. [10,33]). From obvious topological considerations, one then has to deal with at least 3 free parameters to account for the "low" and "high" citation modes, and has to add 2 more (1 per component) – to reflect the curving deviations from the "canonical" power law behavior, if pursuing to accurately mimic the observed frequencies. Given the complexity of the model, however, there does exist a risk that the proposed formula "overfits" the data by reproducing truly random fluctuations (i.e. the noise) along with the underlying function to be approximated. Supporting arguments other than the proximity of the model probability distribution to the observed citation frequencies are, therefore, indispensable.

A function extensively used in psychology and social sciences to analyze proposed models and highlight their important points is the hazard rate $h(t) = \dfrac{f(t)}{1 - F(t)}$ defined via $f(t)$ and $F(t)$, the PDF and the cumulative probability function (CDF) of a distribution, respectively. Figure 3 depicts two hazard functions (normalized by the expectation) for the probability distributions of the processing time $f(\tau | b \equiv 1, \alpha = 1/\mu, \sigma = 1/\sqrt{\lambda})$ (the corresponding density functions (5) are shown in the inset), reconstructed with parameter values estimated from the Physical Review data (Fig. 2, **A**). In the context of this study, the hazard function $h(\tau)$ can be understood as a measure of the depreciation rate of the concept (problem, idea, etc.) associated with a cited paper – a rate, at which a problem is solved or/and becomes uninteresting from the standpoint of citation at time $\tau$, given that the problem has not been solved at an earlier time. It is important to notice that the fitted model unveils two quite characteristic (to human decision-making) patterns in the Wald-distributed processing time: one associated with frequently cited papers (solid line), and another – with papers, which receive few citations (dashed line). Papers of the "repeated citation" pattern were referred to, on average, 11 times more often (as estimated via $E(\beta)$) than papers with the "new citation" processing time. At the same time, it is apparent from the graphs that works of the former group are not depreciated as fast as the ones of the latter do. These findings well accord with our conjecture about the bursty character of the citation process formalized with Eq. (3), and it appears very unlikely that the two patterns in the processing time would be detected using parameters of a merely interpolating form (see Ref. [30]). This, together with the fact that the second data collection (the high energy physics articles) uncovered, as it is implied from the estimated model parameter values given in Fig. 2 (B), a very similar structure of the citation process, allows us to argue that the approximations obtained for the citation statistics are not driven by random oscillations – the model (7) does not overfit the data.

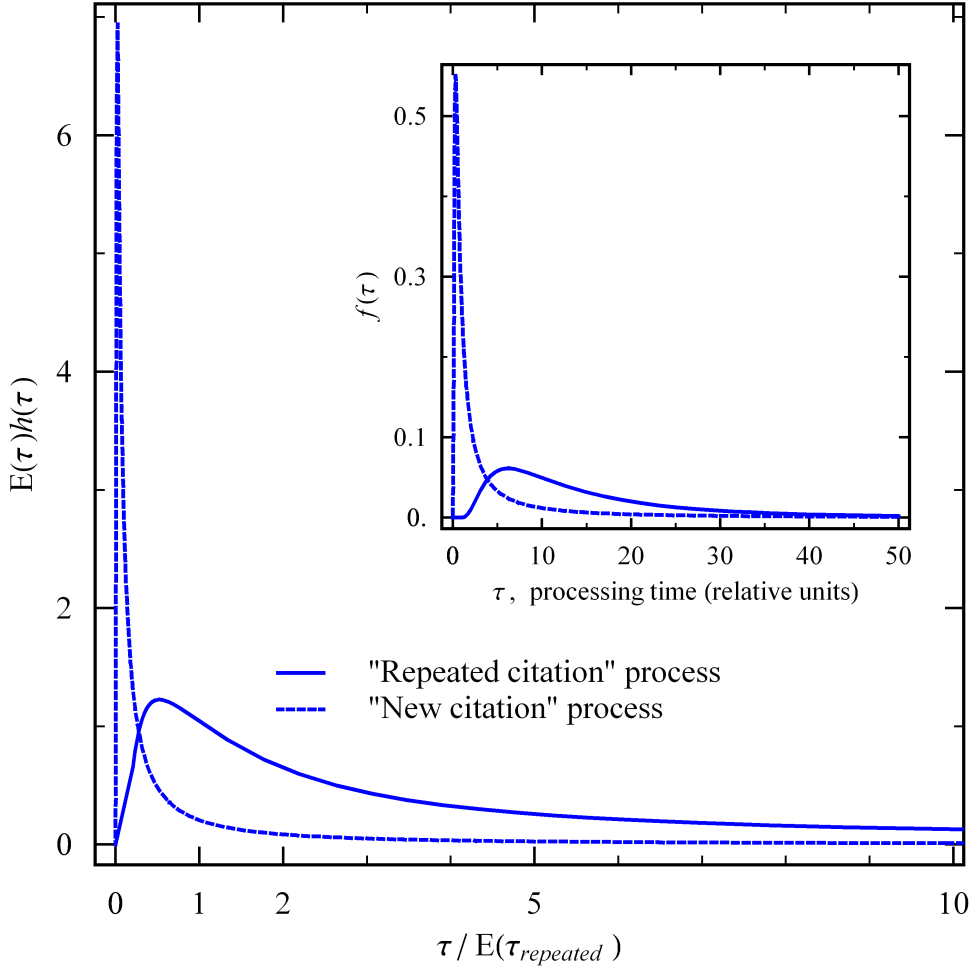

**Fig. 3.** Characteristic patterns in the time devoted to a problem in a research community as revealed by citations in Physical Review D in 1975-1994 (on the axes, E(·) signifies the expectation operator): the reconstructed hazard (main figure) and probability density (inset) functions.

Let us now consider a continuous analog of the PMF (7). For each of its two components, the corresponding PDF can be written as

$$f(x) = \frac{\lambda\sqrt{2x+\lambda} + \mu\sqrt{\lambda}}{\mu(2x+\lambda)^{3/2}} e^{\frac{\lambda - \sqrt{\lambda(2x+\lambda)}}{\mu}}, \qquad (8)$$

where $x > 0$ is the measured stochastic variable, and parameters $\lambda$, $\mu$ are defined in terms of the diffusion process in the same way as their counterparts of the discrete distribution. It is interesting to observe that for large $\mu$ (i.e. for "extremely hard or/and eternally important" problems), Eq. (8) becomes a power law with a fixed exponent $\frac{3}{2}$: $\lim_{\mu \to \infty} f(x) = \sqrt{\lambda}(2x+\lambda)^{-3/2}$. Likewise, for large $\lambda$ (i.e. for "deterministic" problem-solving and/or very big activation effort), it is a pure exponential: $\lim_{\lambda \to \infty} f(x) = \frac{1}{\mu} e^{-\frac{1}{\mu}x}$. As can be seen from Fig. 4 (A), these two limiting distributions would be positioned so as to alone account for the empirical data in the "high" (power law) and "low" (exponential) citation regions. The experimental results displayed in Fig. 2 suggest, however, that since the estimated parameter values of $\lambda$ are by no means even close to being large, the observed citation frequencies could hardly be explained in terms of low-stochasticity orthogonal forms defined with $\lambda \gg 1$ (note also the generally poor fit of the exponential form obtained via MLE). This makes questionable persistent speculations (e.g. as in Ref. [34]) that many human "production processes," including citation and text-based communication, would completely be characterized in terms of an exponential-family distribution. On the other hand, Fig. 4 (A) corroborates that setting a boundary somewhere between 40 and 200 citations could naturally partition the empirical histogram into the "low" and "high" components, as it is from within this range of citation frequencies that the power law would, on its own, approximate the data behavior (see also Fig. 1, inset). Such (ad hoc) subdivisions were proposed in several publications [10,33].

In the context of our study, a more interesting partitioning emerges from discovering conditions when $(1-c)P_2(k) > cP_1(k)$ holds, where $P_1(k)$ stands for the first component of the PMF (7), and $P_2(k)$ – for the second. Numerical solution of this inequality (see Figs. 4, B and C) results in $k \in [4, 118]$ for the data shown in Fig. 2 (A), and $k \in [2, 253]$ – for the data of Fig. 2 (B). The thus detected intervals allow us to speculate about citation frequencies most likely caused by a burst – a sudden boost of attention to a specific work – in the respective research communities (in Figs. 4, B and C, these are the frequencies bounded by the dashed lines). Articles received fewer citations than the minimum in the burst would be considered as "socially not acknowledged" at the moment, while articles with more citations than the maximum in the burst would be nominated as "classic," i.e. works that triggered bursts repeatedly. This interval-based partitioning appears more natural than the traditional single number -based classification of (cited) papers, as it better, in our opinion, agrees with the general intuition of what citation rate would be a sign of prominence and what rate would be routine.

Overall, the asymptotic properties of the developed model are consistent with common-sense expectations: utterly deterministic decision-making contributes little to the "creation of fame" in the citation process, but it may well account for a significant part of the total citation occurrences – for "no-or-few-citation" papers, which make up an absolute majority of published works. In contrast, the phenomenon of highly cited papers may completely be explained by a lasting social attention to the ideas dealt with in such papers, so that the corresponding Wald-distributed processing time $\tau$ with a mean value $\mu$ linearly determines the citation rate. A considerable social interest (and, hence, presumably large $\mu$) is, however, a necessary but not a sufficient condition for an article to be (re-)cited, as Eq. (8) entails that $f(0) \to 0$ only if both $\mu \to \infty$ and $\lambda \to \infty$. Henceforward, we would like to refrain from further speculating about the interplay effects of $\mu$ and $\lambda$, since we strongly believe that analytical properties of any model should be investigated in conjunction with some form of experimental assessment of the model parameters in focus that, though naturally, exceeds the bounds of this particular work.

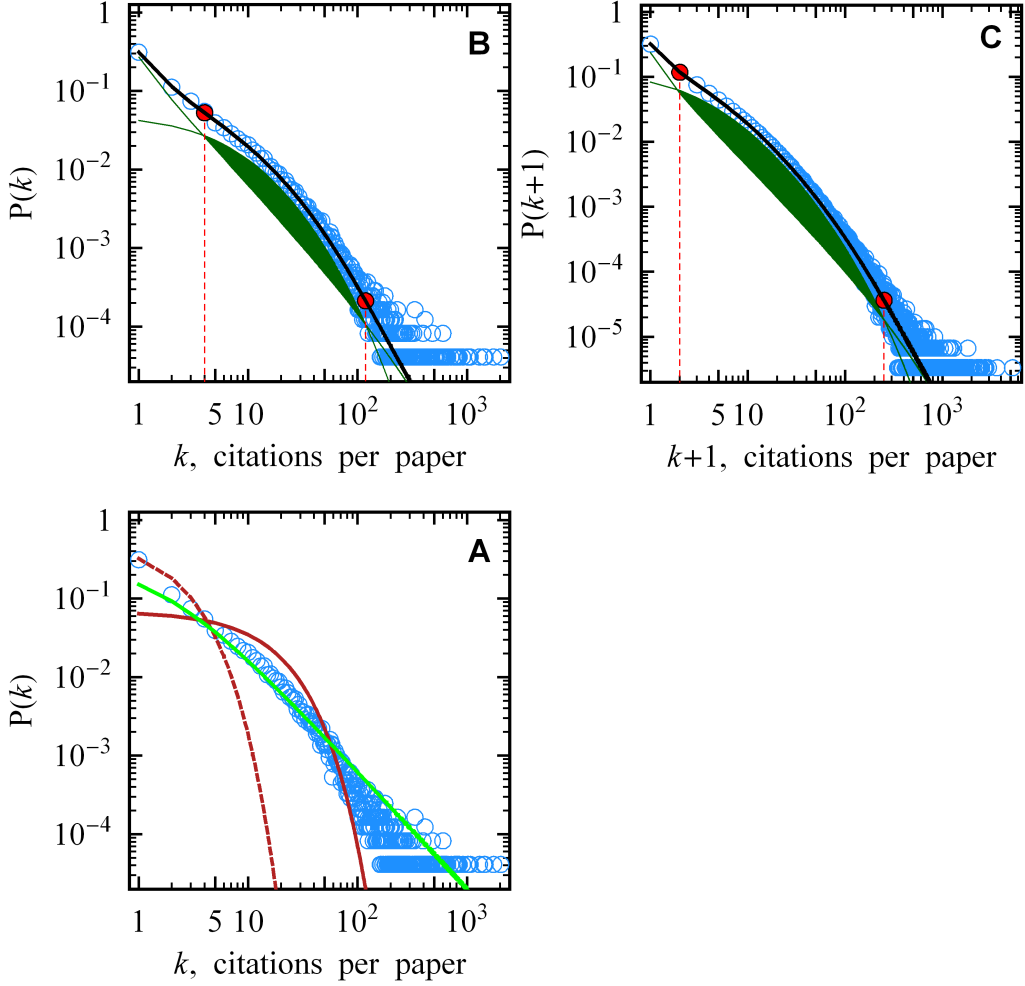

**Fig. 4.** The "low-" and "high- citation mode" partitioning of the empirical data (shown with circles): (A) The Physical Review dataset and the limiting forms of $f(x)$, the distribution function (8) specified as $\lim_{\lambda \to \infty} f(x) = (1/\mu) e^{-x/\mu}$, the exponential (solid dark line – an MLE fit, dashed dark line – a non-linear MSE fit), and $\lim_{\mu \to \infty} f(x) = \sqrt{\lambda}(2x+\lambda)^{-3/2}$, the power law (solid light line, MLE). (B) The solution of the inequality $(1-c)P_2(k) > c P_1(k)$ for the same dataset; the filled region is created by the upper bound $(1-c)P_2(k)$ and the lower bound $c P_1(k)$ curves. (C) The same inequality is solved for the SLAC SPIRES data. (In B and C, the dark solid line shows the fit by model (7) with the parameter values presented in Figs. 2, A and B, respectively.)

As a closing remark of this section, we note that Eq. (8) is reminiscent of the Weibull density function $f(x) = \frac{a}{b}\left(\frac{x}{b}\right)^{a-1} e^{-\left(\frac{x}{b}\right)^a}$ – a probability distribution, various forms of which are typically discussed as alternatives to power law and lognormal models of complex phenomena [35,36,12]. Unlike the Weibull family, however, the form derived in this paper has no readily identifiable cut-off point that would differentiate the exponential and power-law components.

## 5.    Conclusions and future work

The main contribution of the presented study is a possible explanation of the citation process as a communication phenomenon from the positions of cognitive psychology. The proposed model opens a number of new research avenues and analytical prospects. The distribution of "one concept" citation frequencies (2) derived in the beginning of Section 2 implies, due to Bernstein's theorem [37], that there always exists some proper probability distribution function characterizing $\tau$ the time, for which specific publications are cited. By reconstructing the "true" form of $f(\tau)$, the PDF of $\tau$, from observed citation frequencies (e.g. by specializing the very general Beta of the second kind model [38]), one would then acquire a tool to analyze how new ideas propagate in one or another scientific community. The problem-solving "productivity"/information-processing capacity of various communities (or, based on the corresponding citation habits, individuals) would be described in terms of the corresponding hazard functions $h(\tau)$, e.g. in relation to specific topics and/or groups of authors (see Ref. [39] on the use of hazard functions as processing capacity indicators). Besides, $f(\tau)$ the reconstructed probability functions would, on their own, be used as an alternative to the "citation pattern" – a characteristic recently suggested to compare citation practices in different domains [35]. These and many other possibilities in citation analysis would require, however, for the introduced model basic assumptions to hold in a broad context of the involved cognitive and social systems – an issue this given paper does not address. In view of the latter, it appears important to extend the theoretical framework for the case of the so-called "non-extensive systems" [40]: a challenging task would be to find out which of the known (or at least legitimate) entropic forms (see Ref. [41]) is the "best" to contemplate the dynamics of biological and social systems. One may also be interested in looking for a better (in whatever it would mean) model for human decision-making: the chosen parameterization for the Inverse Gaussian distribution and, more generally, the diffusion model used in the presented study may not be the best candidates to analyze the essentially social dynamics of public attention [32].

We would like to conclude this work by words ascribed to a prominent American scholar Henry Louis Mencken, who observed in the distant 1917 that "…[t]here is always an easy solution to every human problem – neat, plausible, and wrong." Our theory may not necessarily correctly or fully explain the "true" mechanisms of citation. It does, however, provide for an accurate description of citation statistics and is in a good agreement with relevant studies in related domains, such as psychology and sociology.

## Acknowledgments

The authors would like to thank Sune Lehmann for sending us the SLAC SPIRES data collection. Useful discussions with Eric W. Cooper and Tove Faber Frandsen are also gratefully acknowledged.